# Intergrain current flow in a randomly oriented polycrystalline SmFeAsO$_{0.85}$ oxypnictide

F. Kametani, P. Li, D. Abraimov, A. A. Polyanskii, A. Yamamoto, J. Jiang, E. E. Hellstrom, A. Gurevich and D. C. Larbalestier

Applied Superconductivity Center, National High Magnetic Field Laboratory, Florida State University, Tallahassee FL 32310, USA

Z. A. Ren, J. Yang, X. L. Dong, W. Lu and Z. X. Zhao

National Laboratory for Superconductivity, Institute of Physics and Beijing National Laboratory for Condensed Matter Physics, Chinese Academy of Science, P.O. Box 603, Beijing 100190, P.R. China

## Abstract

We report a direct current transport study of the local intergrain connections in a polycrystalline SmFeAsO$_{0.85}$ (Sm1111) bulk, for which we earlier estimated significant intergranular critical current density $J_c$. Our combined low temperature laser scanning microscopy (LTLSM) and scanning electron microscopy observations revealed only few grain-to-grain transport current paths, most of which switched off when a magnetic field was applied. These regions typically occur where current crosses Fe-As, which is a normal-metal wetting-phase that surrounds Sm1111 grains, producing a dense array of superconducting-normal-superconducting contacts. Our study points out the need to reduce the amount of grain boundary-wetting Fe-As phase, as well as the crack density within pnictide grains, as these defects produce a multiply connected current-blocking network.

Many aspects of the superconducting behavior of the broad class of new superconductor Fe-oxypnictides have been rapidly explored since they were discovered[1-13]. In particular, the rare earth Fe-oxypnictides (RE1111) with typical composition REFeAsO$_{1-x}$F$_x$ have drawn great interest because of their high transition temperatures Tc[4-7]. They also have a very high upper critical field H$_{c2}$ of ~65 T even in La1111 with Tc ~26 K[8], while H$_{c2}$ over 200 T was deduced for Sm- and Nd1111[9]. However the absence of significant transport currents in polycrystalline samples has also raised the concern that there is a significant depression of the superconducting order parameter at grain boundaries (GB)[10-12], resulting in randomly oriented grain boundaries weak links. Such behavior would be similar to that of the cuprates, where this GB depression is very significant, causing the exponential decrease of the GB critical current density J$_{cGB}$, as the grain misorientation angle $\theta$ increases above 3-5°[14-17]. As a result, J$_{cGB}$ becomes much lower than J$_c$ sustained by vortex pinning within the grains. To address this point, Yamamoto et al[13] performed remnant magnetization and magneto optical studies of polycrystalline Nd- and Sm1111 and uncovered evidence for a significant intergrain current density (global J$_c$) up to ~4000 A/cm$^2$ at 4.2 K some 10 times higher than in random polycrystalline RE123 cuprates. However the intergrain and intragrain current densities had different temperature dependences and differed by 3 orders of magnitude, leaving open the possibility of an intrinsic GB blocking effect.

To address the nature of the active current paths in such polycrystals, we then combined microstructural and magneto optical (MO) images of the same Nd- and Sm1111 bulk samples made under high pressure[18]. Using back scatter electron scanning electron microscopy (BSE-SEM) to show both the grain structure and its local orientation, we were able to see that almost all high $J_c$ spots observed in the MO images came from individual RE1111 grains. No convincing cases of randomly misoriented grain pairs with a strong coupling were seen by MO. Moreover, the BSE-SEM images revealed that even the best Sm1111 bulk with a deduced global $J_c$ of ~4000 A/cm$^2$ at 4.2K had a multi-phase microstructure in which non-superconducting Fe-As and $RE_2O_3$ occupied at least three quarters of the RE1111 grain boundaries, making the active current path certainly much smaller than the geometrical cross-section of the sample.

To further resolve this issue, we have now combined low temperature laser scanning microscopy (LTLSM) and SEM studies to reveal the active local current paths. Depending on specimen thickness, the LTLSM has a spatial resolution down to 1-2 μm for the electric fields created by the supercurrent. External magnetic fields of up to 5 T can be applied, thus allowing comparison of local electric fields in the self field and the high field state. The combined images allowed us to correlate current and electric field distributions with microstructure on a well-polished ~20 μm thick sample cut from the same bulk of Sm1111 we reported on earlier[18]. In LTLSM measurements, a superconducting sample is cooled below $T_c$ while biased with transport current $I_{bias}$ so that a measurable voltage occurs. A laser beam 1-2 μm in diameter is scanned over the sample surface[19-21], producing a 2-3 μm diameter hotspot where the local temperature is increased by 1~2 K. A local decrease in $I_c$ around the hotspot results in a small increase in voltage change dV across the whole sample if the $I_c$ of locally heated cross section becomes smaller than the $I_{bias}$. The dV response of the LTLSM measured at each point is proportional to the local electric field[22]. In non-uniform superconductors, a local dV response is larger near blocking defects due to their preferential focusing of the supercurrent into channels[22,23]. Here we show that we can clearly distinguish field-sensitive and we believe Josephson-coupled paths across grain boundaries wetted by Fe-As phase, from constricted current paths produced by local cracking which are common in these Sm1111 samples.

The polycrystalline $SmFeAsO_{0.85}$ (Sm1111) bulk sample was synthesized by solid state reaction under high pressure. Pre-sintered SmAs, Fe and $Fe_2O_3$ powders were mixed together in stoichiometric ratio as noted elsewhere[6]. Even though the microstructure is far from single phase[18], the sample has a sharp magnetic $T_c$ transition and significant global Jc[13]. In fact, samples of apparently similar quality as judged by x-ray and scanning electron microscopy examination showed much lower intergrain current density[24, 25, 26].

Our Sm1111 bulk sample was polished down to ~ 20 μm thickness after being glued to a sapphire substrate with M-BOND, and finished with 50 nm colloidal silica, so that the LTLSM could detect as many active current paths as possible. LTLSM maps were scanned with a tightly focused red beam ($\lambda$ = 639 nm) delivered by single mode fiber from laser diode (maximum power 17.5 mW ) and TTL modulated at 102 kHz. Images were taken at 39K with a constant bias current ($I_{bias}$>$I_c$) that kept the bridge voltage at 100 μV (±2 μV) independent of field. Simultaneously with the dV signal, we recorded the reflected laser beam intensity so as to have a photo image to correlate the dV maps with subsequent SEM images taken in a Carl Zeiss 1540EsB.

Fig. 1a shows the BSE-SEM image and LTLSM dissipation maps on the 150 µm wide, 300 µm long and ~20 µm thick Sm1111 bridge. The BSE-SEM of Fig.1a and Fig. 2 shows dark grey FeAs, white $Sm_2O_3$, and platelet light grey Sm1111 grains[18]. Fig. 1b, the self-field LTLSM map, shows a highly inhomogeneous distribution of ~20 dissipation spots where the local electric field $E(x,y)$ is increased by the current being squeezed into narrow paths by current-blocking defects[22,23]. These LTLSM patterns clearly show that only a few current paths are active. But application of even 0.1 T shrinks the dissipation spot density dramatically (Fig. 1c), and after applying 5 T (Fig. 1d), even fewer are visible. Thus even small fields break most of the intergranular superconducting paths. For further discussion, we compare switch-off spots A, B and field-resistant spot C from Fig. 1.

The SEM images of Figs. 2a-2c show significant microstructural differences between spots A and B, and C, respectively. Insulating $Sm_2O_3$ has a small surface to volume ratio and is mostly located within Sm1111 grains, so it has the smallest effect on current transport. By contrast, the dark grey Fe-As phase wets many GBs, thus interrupting grain-to-grain supercurrent paths, which are further degraded by extensive cracking, sometimes at GBs (the black appearing lines) and sometimes within grains.

At switch-off spot A (Fig. 2a), a crack F on the upper side and a grain of $Sm_2O_3$ and a large FeAs region G force current to cross the GB (H) which contains a thin Fe-As layer, producing the dissipation spot seen in the overlay of Fig. 2d. At switch off spot B of Fig. 2b and 2e, the current is channeled by cracks, Fe-As and $Sm_2O_3$ into a narrow passage crossing Fe-As regions too. By contrast, as shown in Fig. 2c and 2f, spot C which remains dissipative even in 5 T field has its peak dissipation within a single grain at a constriction provided by two almost orthogonal sets of cracks that squeeze the current between the two diagonal cracks.

Fig. 3 shows a strongly field-dependent global current density $J_c$ of the bridge suggestive of Josephson- coupled connections across a normal (N) metal contact such as the wetting Fe-As phase. $J_c$ falls from ~470 A/cm$^2$ to ~30 A/cm$^2$ as 0.1 T is applied. The electric field versus current density (E-J) curves in fields of 0-5 T (inset of Fig. 3) are indeed supercurrent curves and thus are consistent with the local $J_c$ being much higher than the whole-cross-section average of 30 A/cm$^2$. These E-J curves do support our earlier arguments that there is a true supercurrent characteristic of a vortex pinning current that is crossing GBs. The almost field independent $J_c$ between 1 and 5 T is consistent with the earlier observation that the irreversibility field $H_{irr}$(39 K) for this sample is ~25 T[9]. The self field transport $J_c$ ~470 A/cm$^2$ at 39 K shows a good agreement with the whole-sample global $J_c$ of ~600 A/cm$^2$ estimated earlier by remnant magnetization and MO analyeses[13], although the transport $J_c$ is slightly suppressed presumably due to extensive cracks at grain connections caused by thermal stress during the sample preparation and measurements.

In summary, we combined microstructural and LTLSM current dissipation studies to investigate local supercurrent paths on a carefully ground ~20 μm thin sample made from a Sm1111 bulk The present data show how compromised the connectivity of present polycrystals is, mostly by cracks and a grain boundary Fe-As phase. From Fig.2 we can hardly explain the poor connectivity of the polycrystal as being due to any intrinsic property of the Sm1111 grains, since clearly the Fe-As and the cracks form an obstructive 3D network through which current percolates. Moreover we cannot ascribe the poor connectivity to any intrinsic properties of Sm1111 grain boundaries, since there are too few regions where the GBs are not covered by wetting Fe-As. However, the residual, albeit small high field supercurrent does indicate that there must be some strongly coupled Sm1111 GBs, but whether they are rare low angle connections, of which very few were seen in our earlier study[18], or more random, higher angle connections is impossible to say at this time. Clearly bicrystal measurements of $J_c$ in good thin films would be very valuable in taking our understanding to a deeper level.

This work was supported by the NSF Cooperative Agreement DMR-0084173 and by the State of Florida, and by AFOSR under Grant No. FA9550-06-1-0474.

Fig.1 (a) Back scattering electron SEM image of whole Sm1111 bridge. Sm1111, FeAs and $Sm_2O_3$ are shown as grey, black and white contrasts respectively. (b-d) Supercurrent dissipation map recorded at 39K and self field with bias current density $J_{bias}$ of 1111 $A/cm^2$, 0.1T with 348 $A/cm^2$ and 5T with 198 $A/cm^2$, respectively. The $I_{bias}$ direction is also shown by an arrow. Most of dissipation spots disappear in magnetic fields, while the strongest spot remains, as typically shown within the black oval. Typical field-dependent spots A, B and an in-field dissipation spot C are shown in details in Fig.2.

Fig.2 (a-c) Higher magnification SEM images showing respective microstructural details of field-sensitive switch-off spots A, B, and a field-insensitive dissipation spot C in Fig.2. There are 2nd phases of $Sm_2O_3$ and FeAs with white and dark grey contrast in addition to Sm1111 platelet grains. Cracks with dark line contrasts are also seen. (d-f) The self-field dissipation spots are superimposed on the SEM images at right. Deeper red color shows the areas with stronger dissipation where the supercurrent is preferentially focused.

Fig.3 Field dependence of the critical current density Jc at 39K showing that Jc sharply declines when small fields are applied and also that there is little dependence of Jc for H between 0.5 and 5 T. Inset shows electric field versus current density curves at 39K in fields of 0-5T

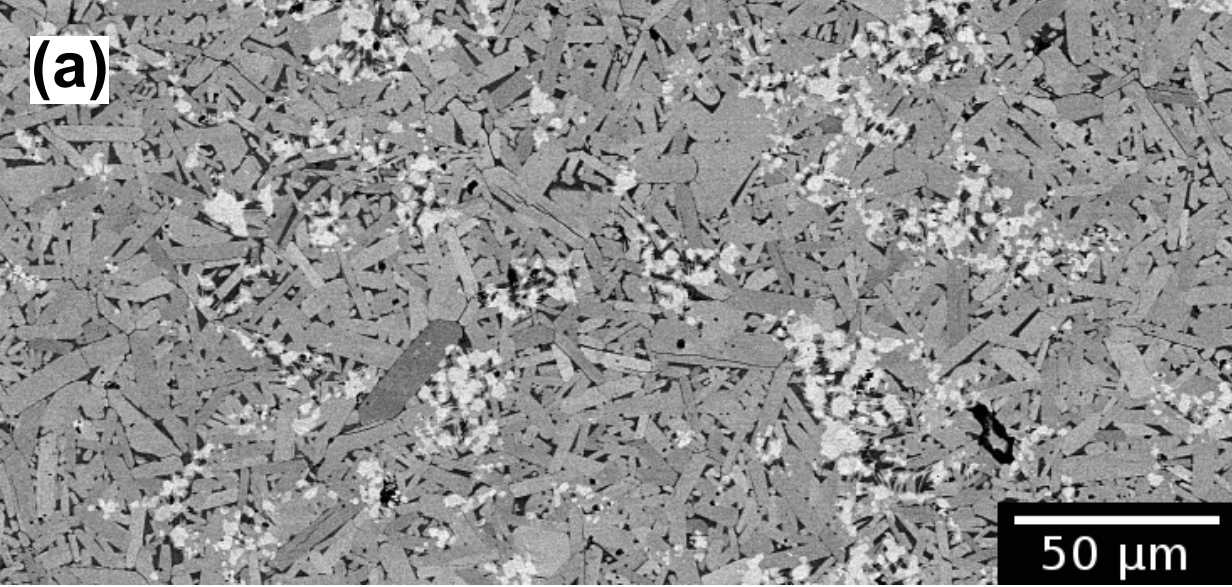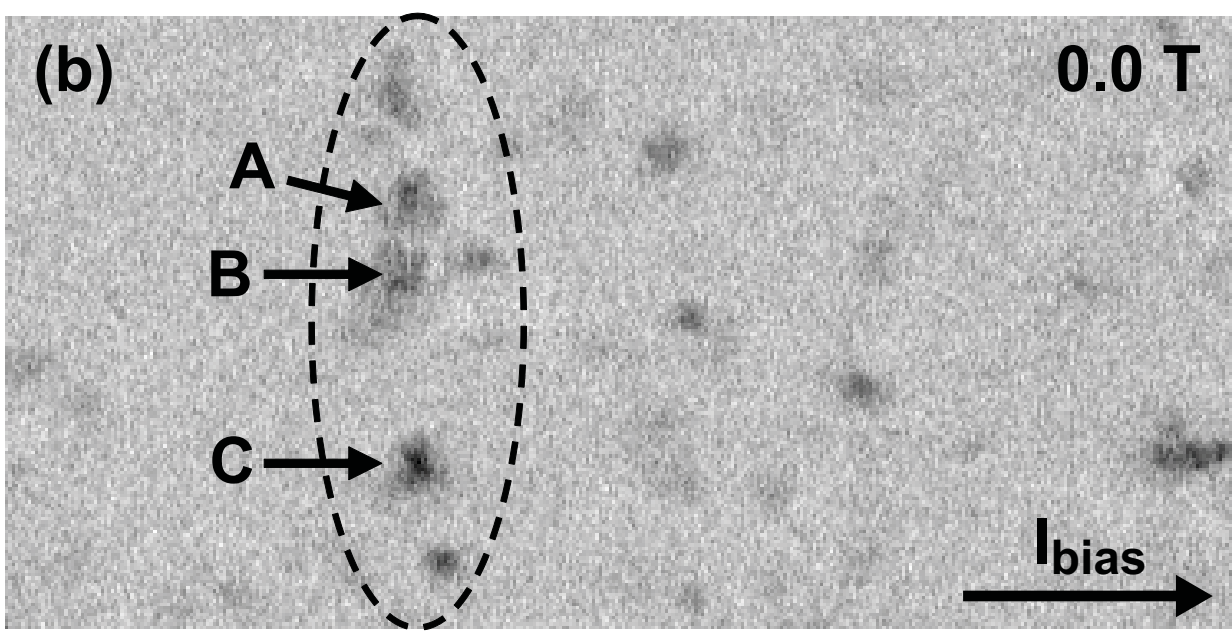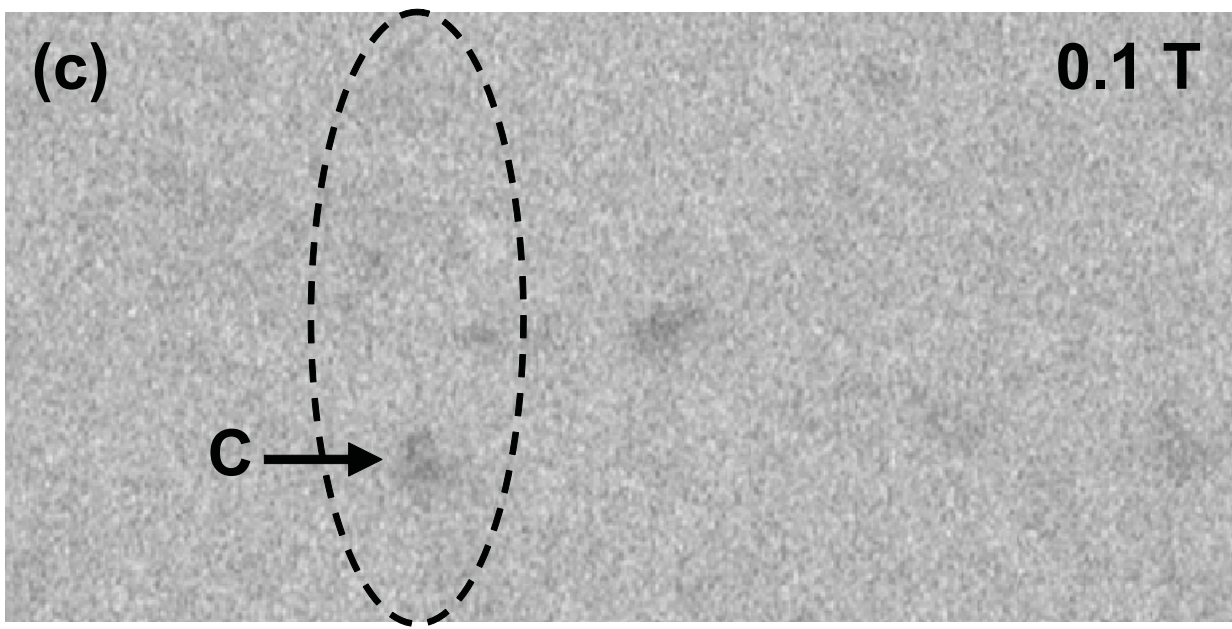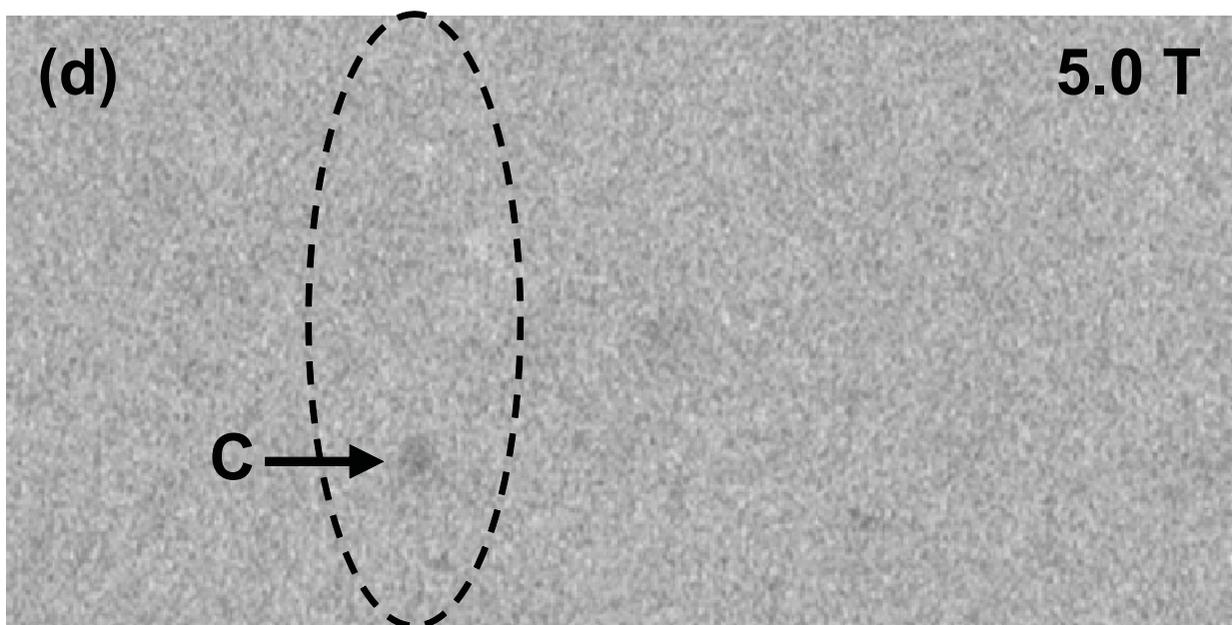

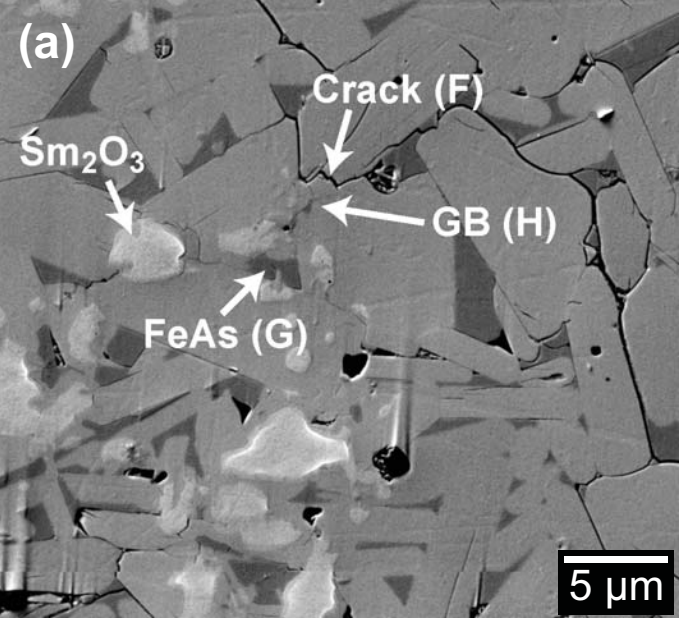
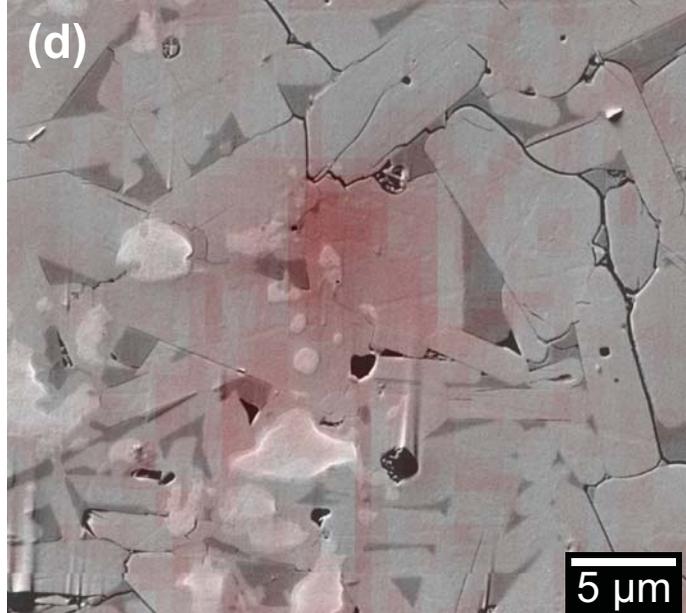
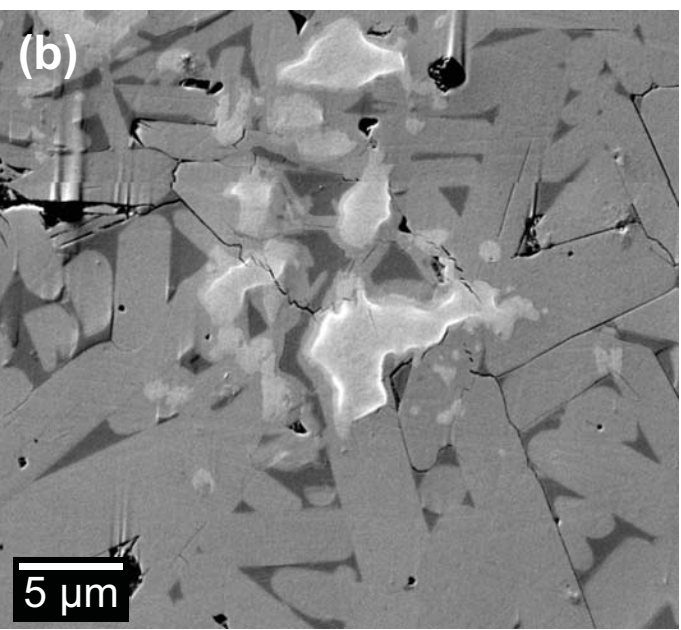
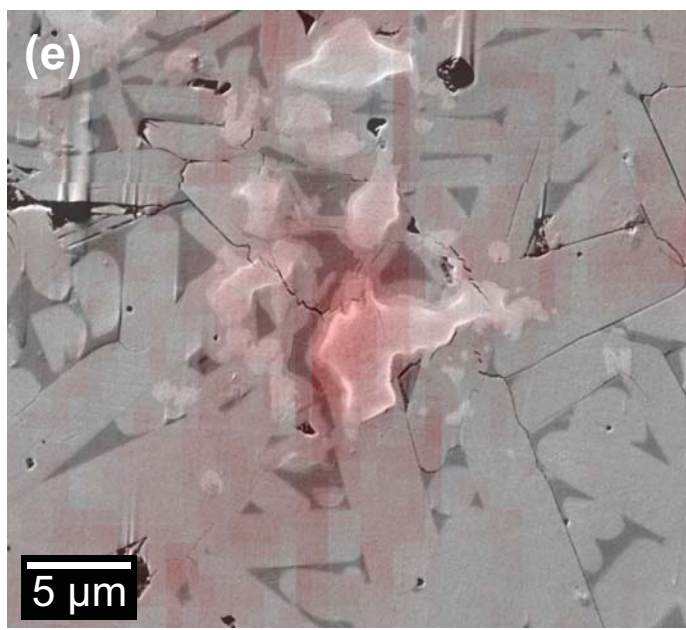
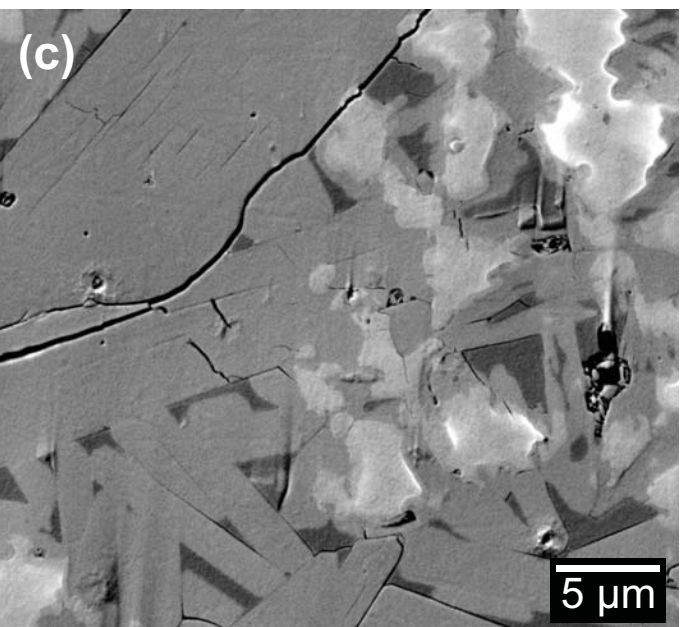
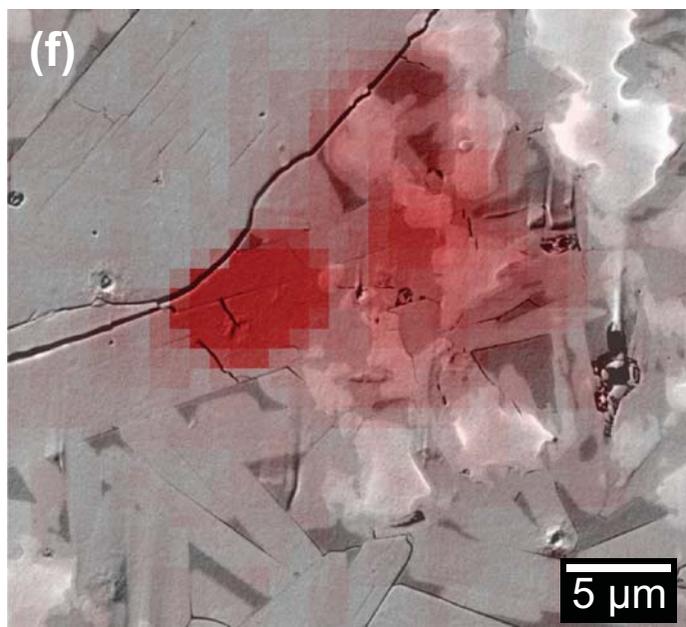

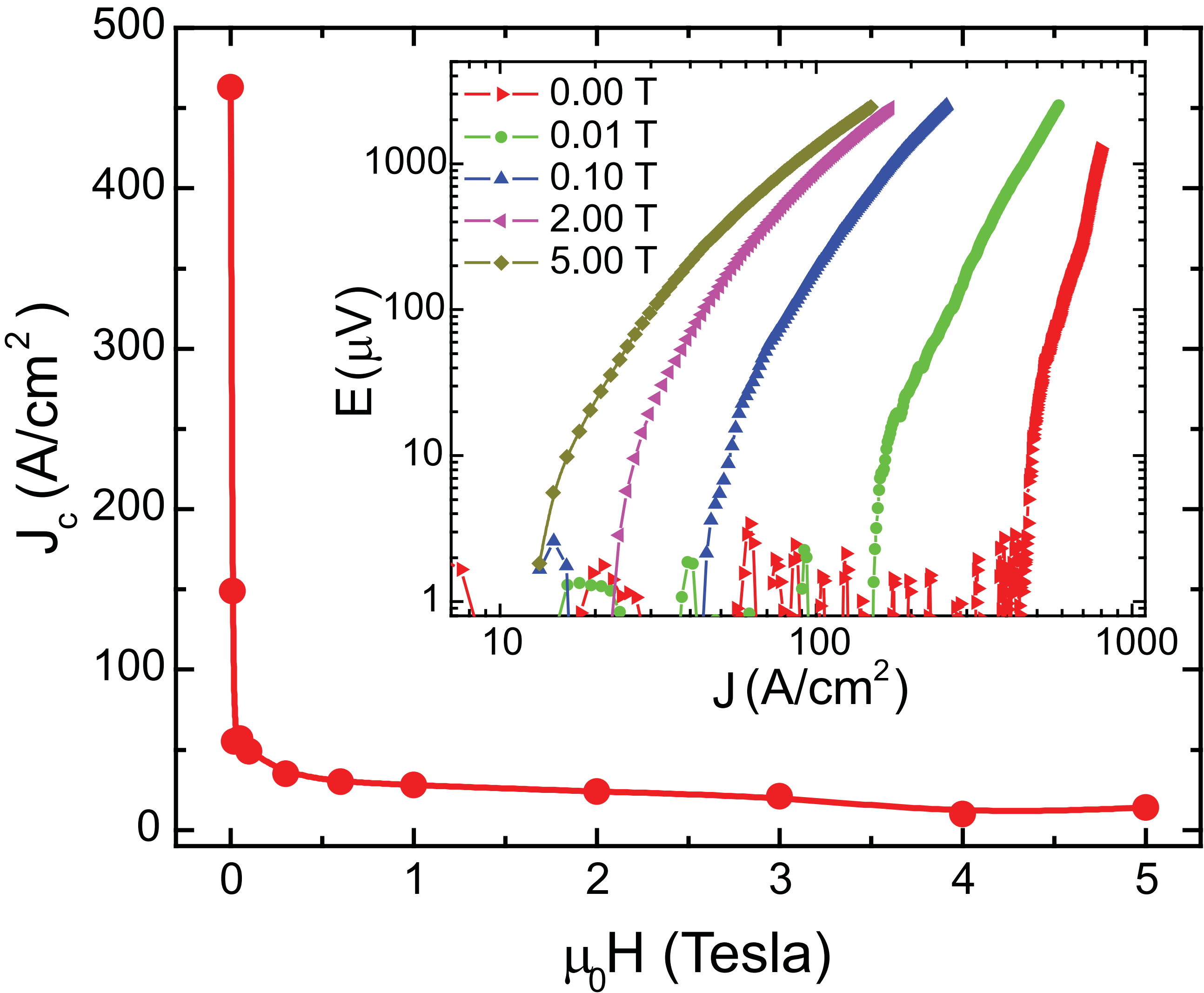